# Temporal-multiplexing interferometry applied to co-phased profilometry

**MANUEL SERVIN,\* MOISES PADILLA, AND GUILLERMO GARNICA**

*Centro de Investigaciones en Optica A.C., Loma del Bosque 115, 37150 Leon, Guanajuato, Mexico.*
*\*mservin@cio.mx*
*https://www.cio.mx*

**Abstract:** Fringe-projection profilometry with 1 camera and 1 fringe-projector is a well-known and widely used technique in optical metrology. Spatial-frequency multiplexing interferometry with several spatial-carriers having non-overlapping spatial-spectra is well known and productive in optical metrology. In this paper we propose temporal-multiplexing phase-shifting interferometry applied to profilometry. That is, instead of having fringe-patterns with well separated spatial-spectra, we propose instead to separate the fringe information in the temporal-spectra. In other words, we may have overlapping spatial-spectra, but separated in the temporal-spectra by frequency multiplexing. Using 1-camera and several fringe-projectors one minimizes the object shadows and specular reflections from the digitizing solid. Temporal multiplexing profilometry allows us to illuminate the object from several projectors turned-on simultaneously. In previous phase-shifting co-phased profilometry, the projectors were turned-on and off sequentially. As seen in this work temporal-multiplexing allow us to demodulate the several fringe-patterns without crosstalk from other simultaneously projected fringes. This is entirely analogous to having several television stations broadcasting simultaneously, each TV-transmitter having its own broadcasting frequency. A given TV-receiver tunes into a single TV-station and filter-out all other broadcasters. Following this analogy, each fringe-projector must have its own temporal broadcasting frequency to remain well separated from all other projectors in the time-spectra domain. In addition to the general theory presented, we assess its feasibility with experimental results.

## 1. Introduction

Fringe projection profilometry applied to digitize three dimensional (3D) objects has been used for a long time [1]. A basic fringe projection profilometer consists of a multimedia digital projector, a CCD digital camera to capture the phase-modulated linear fringes and a computer to phase-demodulate the fringe patterns [1]. The phase sensitivity of a fringe projection profilometer is proportional to the camera-projector sensitivity angle multiplied by the spatial frequency of the fringes [1]. If just a single carrier-frequency fringe pattern is digitalized, it can be phase-demodulated using the Fourier technique [1,2]. On the other hand, when higher resolution digitization is required, we need to process at least three phase-shifted fringe patterns using phase-shifting algorithms (PSAs) [1,2]. The introduction of the frequency transfer function (FTF) formalism for PSAs design have been a significant advance in phase-shifting phase-demodulation [2,3,5,6,7]. However new in temporal interferometry, the FTF formalism has been used in telecommunications engineering for at least 60 years [4]. The inverse Fourier transform of the FTF give us the impulse response of the filter and from it one can easily find the PSA which filters the desired analytic signal [2,3]. The analytic signal preserves the amplitude and the phase of the temporal interferograms in a complex-valued signal; this is paramount in co-phased profilometry [2,3,5-7]. As briefly reviewed in this work, in co-phased profilometry, *N* fringe-patterns are projected towards the digitizing object from different directions having equal sensitivity angles and equal spatial-carrier [5,7]. In this way, the complex-valued signals estimated from each fringe projection are co-phased. Having several co-phased analytic signals one simply adds them to obtain the solid's digitalization with minimized shadows and specular reflections [5-7]. In addition to this, co-phased profilometry coupled to 2-steps temporal-unwrapping has substantially improved fringe-projection profilometry [7] with respect to the previous state of the art [8-22].

In contrast, spatial frequency multiplexing has been used in interferometry and profilometry for years [8]. In spatial frequency-multiplexing, several linear fringe-patterns, each one having non-overlapping spatial-spectra, are projected into the object. The spatial spectral-lobes must be well separated (non overlapped), to phase demodulate them using spatial quadrature-filters [11-22]. In temporal-multiplexing however each projected fringe-pattern must *oscillates* at a unique temporal frequency, in this way all fringe-projectors can be turned-on simultaneously. As shown here, in temporal-multiplexing one must use as many temporal-frequencies as fringe-projectors simultaneously illuminate the object under analysis.

This is entirely analogous to public radio broadcasting television (TV). Several TV-stations may transmit simultaneously, but each one must have its own broadcasting frequency [4]. Therefore, each TV-receiver tunes into a single TV-station flatly rejecting all other TV-stations. Even though frequency multiplexing in telecommunications engineering is more than 100 years old [4], as far as we know temporal multiplexing has not been explored yet in phase-shifting profilometry. In this paper, the digitizing 3D object is simultaneously illuminated from several fringe-projectors simultaneously. Each fringe-projector must have its own temporal phase-shifting step (broadcasting frequency). In this way, one may design temporal multiplexed PSAs capable of tuning-in at different temporal frequencies. Therefore the main objective of this paper is the synthesis of temporal-multiplexed PSAs for fringe projection profilometry when several projectors illuminate the object simultaneously from different directions. Previous non-multiplexed phase-shifting algorithms (PSAs) were designed to operate at a single temporal-frequency [2,3]. But thanks to the FTF formalism [2], we can now easily design PSAs which operate at several temporal-frequencies (several phase-steps) simultaneously. This generates new interferometric phase-demodulation possibilities for designing temporal-multiplexed phase-shifting algorithms (PSAs).

**2. Mathematical model for 2-temporal-frequencies multiplexed fringes**

The digitizing spherical metallic solid used in the experiments reported herein is shown in Fig. 1(a). The experimental set-up for co-phased profilometry using 2-projectors and 1-camera is shown in Fig. 1(b).

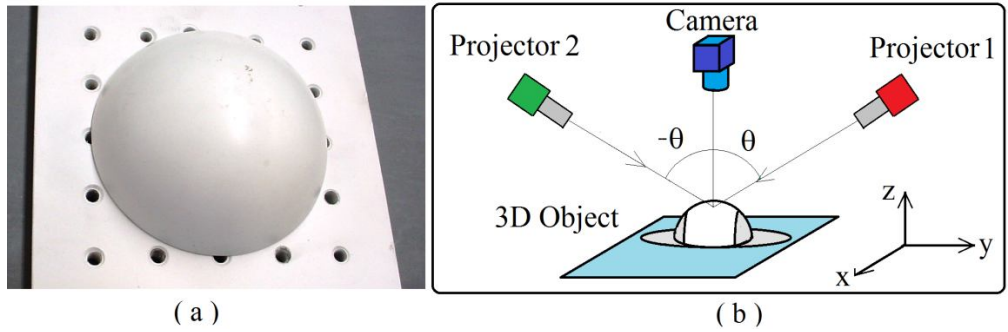

Fig. 1 Panel (a) shows the spheric metallic solid being digitized; this is a regular photography of the object, not a digital 3D rendering. Panel (b) shows the co-phased profilometer composed by 2-projectors and 1-camera. In (b) the shadows of projector-1 are covered by projector-2 and vice-versa. The phase-sensitivity angle of this profilometer is $\theta$.

The digitized phase-modulated fringes for standard 1-camera, 1-projector is given by

$$I_1(x,y,t) = a(x,y) + b(x,y)\cos\left[\, g\,h(x,y) + u_0 x + \omega_0 t \,\right]; \quad g = u_0 \tan(\theta); \quad t \in \mathbb{R}. \quad (1)$$

This fringe image is phase-modulated by $g\,h(x,y)$; the temporal carrier is $\omega_0$; the spatial carrier is $u_0 x$, and $\theta$ is the sensitivity angle. The background illumination is $a(x,y)$ and the fringes' contrast is $b(x,y)$. Finally the digitizing solid is $z = h(x,y)$.

With 2-temporal-frequencies and using the set-up in Fig. 1(b) composed by 1-camera and 2 simultaneous fringe-projectors our previous mathematical model now modifies to,

$$I(x,y,t) = a + b_1 \cos\left[g\,h(x,y) + u_0 x + \omega_1 t\right] + b_2 \cos\left[-g\,h(x,y) + u_0 x + \omega_2 t\right]. \quad (2)$$

The 2 fringe-projectors are multiplexed by two temporal-frequencies $(\omega_1, \omega_2)$. This is shown in Fig. 2. The red trace represents the temporal fringes $b_1 \cos\left[g\,h(x,y) + u_0 x + \omega_1 t\right]$, while the blue trace represents the temporal fringes $b_2 \cos[-g\,h(x,y)+u_0 x+\omega_2 t]$. Finally the lower trace represents their sum $I(x,y,t)$ in Eq. (2).

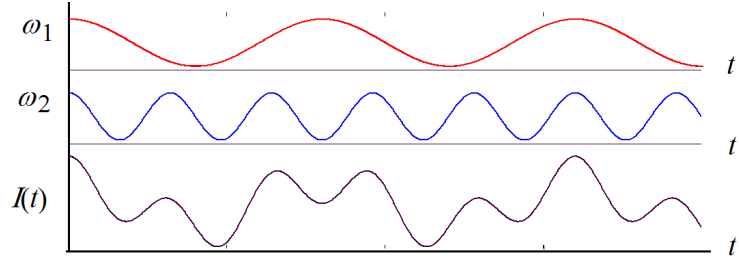

Fig.2 Interferometric fringes at temporal frequencies $\omega_1$ (in red) and $\omega_2$ (in blue). Their temporal multiplexed fringe sum $I(x,y,t)$ in Eq, (2) is shown in black.

The fringes modulated by $b_1(x,y)$ displace in the $x$-direction at a phase-velocity of $(\omega_1)$ radians/image, while the fringes modulated by $b_2(x,y)$ move at a phase-velocity of $(\omega_2)$ radians/image.

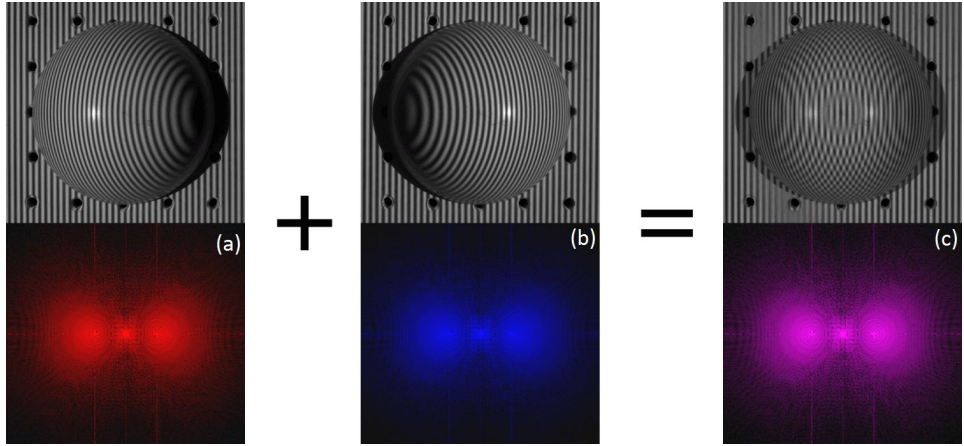

Fig. 3. Fringe-patterns projected over the metallic solid by: (a) the left-side projector only. (b) the right-side projector only. And in panel (c) both projectors simultaneously turned-on. The lower row shows the spatial Fourier spectra of the upper images. The red (left) and blue (right) spectra fully overlap in panel (c) red+blue=purple spectra; precluding their phase-demodulation in the spatial-spectra domain.

Taking the temporal Fourier transform, the multiplexed fringes with amplitudes $b_1/2$ and $b_2/2$ are now well-separated whenever $\omega_1 \neq \omega_2$ (see Fig. 4):

$$I(x,y,\omega) = a\delta(\omega) + \frac{b_1}{2}\left[e^{i(g\,h+u_0 x)}\delta(\omega-\omega_1)+e^{-i(g\,h+u_0 x)}\delta(\omega+\omega_1)\right] + \\ \frac{b_2}{2}\left[e^{i(-g\,h+u_0 x)}\delta(\omega-\omega_2)+e^{-i(-g\,h+u_0 x)}\delta(\omega+\omega_2)\right]. \quad (3)$$

This Fourier time-frequency spectrum is graphically represented in Fig. 4.

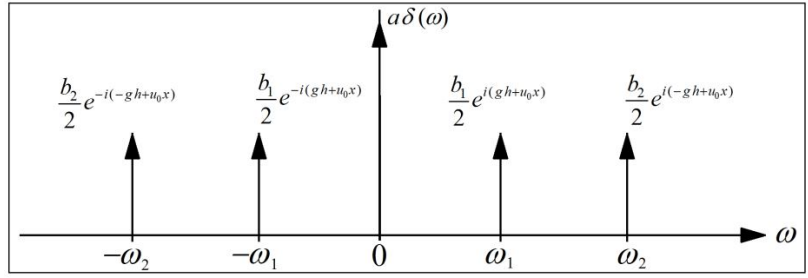

Fig. 4. Fourier spectrum for 2-temporal-multiplexed simultaneous fringe-projections using 2-temporal-frequencies $\omega_1$ and $\omega_2$. The vertical arrows represent Dirac deltas.

Our objective is to estimate the two analytic signals at frequencies $(\omega_1, \omega_2)$. To this end, we need to filter-out the signals at $\omega = \{-\omega_2, -\omega_1, 0, \omega_2\}$ in order to keep $(b_1/2)e^{i(gh+u_0 x)}\delta(\omega-\omega_1)$ at $\omega = \omega_1$. Conversely, we have to filter-out the signals at $\omega = \{-\omega_2, -\omega_1, 0, \omega_1\}$ to keep $(b_2/2)e^{i(-gh+u_0 x)}\delta(\omega-\omega_2)$ at $\omega = \omega_2$. Therefore we need to design 2-PSAs, one tuned at $\omega = \omega_1$ and another one tuned at $\omega = \omega_2$.

## 3. FTF design for 2-PSAs using 5-step time-multiplexed fringe-patterns

Here we use two temporal frequencies namely: $\omega_5 = 2\pi/5$ and $2\omega_5 = 4\pi/5$. As Fig. 5 shows, using the FTF formalism [2,3,7] we need at least four first-order zeroes. One FTF must be tuned at $\omega_5 = 2\pi/5$ and the other one at $2\omega_5$, these are,

$$H_5(\omega) = \left(1 - e^{i\omega}\right)\left[1 - e^{i(\omega-\omega_5)}\right]\left[1 - e^{i(\omega-2\omega_5)}\right]\left[1 - e^{i(\omega-3\omega_5)}\right]; \tag{4}$$

$$H_5(\omega - \omega_5) = \left[1 - e^{i(\omega-\omega_5)}\right]\left[1 - e^{i(\omega-2\omega_5)}\right]\left[1 - e^{i(\omega-3\omega_5)}\right]\left[1 - e^{i(\omega-4\omega_5)}\right]. \tag{5}$$

Figure 5 shows that $H_5(\omega)$ is tuned to $(b_1/2)H_5(\omega_5)e^{i(gh+u_0 x)}\delta(\omega-\omega_5)$, while $H_5(\omega-\omega_5)$ is tuned to $(b_2/2)H_5(\omega_5)e^{i(-gh+u_0 x)}\delta(\omega-2\omega_5)$.

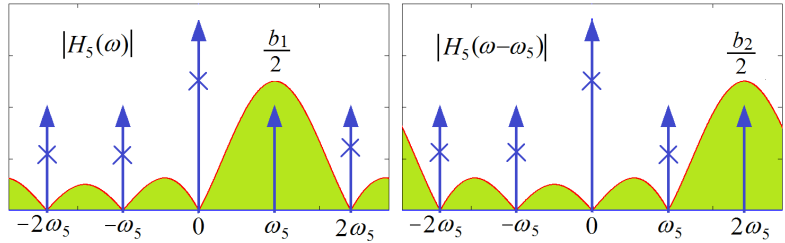

Fig. 5. Fringes' spectrum and FTFs of 5-temporal-multiplexed PSAs tuned at $\omega_5$ and $2\omega_5$.

To find the impulse responses of $H_5(\omega)$ and $H_5(\omega-\omega_5)$, we take their inverse Fourier transforms to obtain,

$$F^{-1}[H_5(\omega_5)] = \delta(t) + \delta(t-1)e^{i\omega_5} + \delta(t-2)e^{2i\omega_5} + \delta(t-3)e^{3i\omega_5} + \delta(t-4)e^{4i\omega_5}, \tag{6}$$

$$F^{-1}[H_5(\omega-\omega_5)] = \delta(t) + \delta(t-1)e^{i2\omega_5} + \delta(t-2)e^{4i\omega_5} + \delta(t-3)e^{6i\omega_5} + \delta(t-4)e^{8i\omega_5}. \tag{7}$$

With $\omega_5 = 2\pi/5$. Therefore the two temporal-multiplexed PSAs are given by [2]:

$$\frac{b_1}{2}H_5(\omega_5)e^{i[gh(x,y)+u_0 x]} = I(0) + I(1)e^{i\omega_5} + I(2)e^{2i\omega_5} + I(3)e^{3i\omega_5} + I(4)e^{4i\omega_5}; \tag{8}$$

$$\frac{b_2}{2}H_5(\omega_5)e^{i[-gh(x,y)+u_0x]} = I(0)+I(1)e^{i2\omega_5}+I(2)e^{4i\omega_5}+I(3)e^{6i\omega_5}+I(4)e^{8i\omega_5}. \qquad (9)$$

The spatial coordinates $(x,y)$ were omitted for clarity. Please note that we are using the same fringe data $\{I(0), I(1),\ldots,I(4)\}$ for both 5-step PSAs tuned at $\omega_5 = 2\pi/5$ and $2\omega_5$. We then remove their spatial carriers $\exp(iu_0x)$ and finally add them (co-phased) as,

$$\frac{H_5(\omega_5)}{2}\left\{b_1e^{i[gh(x,y)+u_0x]}e^{-iu_0x}+b_2e^{i[gh(x,y)-u_0x]}e^{iu_0x}\right\}=\frac{H_5(\omega_5)}{2}\{b_1+b_2\}e^{igh(x,y)}. \qquad (10)$$

We may also that both $H_5(\omega)$ and $H_5(\omega-\omega_5)$ have the same signal-to-noise (S/N) power-ratio gain [2,3] with respect to the raw fringe data, that is,

$$\frac{|H_5(\omega_5)|^2}{\frac{1}{2\pi}\int_{-\pi}^{\pi}|H_5(\omega)|^2 d\omega}=5 \quad ; \quad \frac{|H_5(\omega_5)|^2}{\frac{1}{2\pi}\int_{-\pi}^{\pi}|H_5(\omega-\omega_5)|^2 d\omega}=5. \qquad (11)$$

Without temporal-multiplexing, one would have needed 10-phase-shifted fringes, 5 for each projector sequentially turned-on and off, obtaining the same signal-to-noise (S/N) power-ratio gain with respect to the raw-data fringes (Eq. (2)) [2].

### 4. Experiment with 2 projectors and 2 temporal-frequencies multiplexing

Here we show 2 simultaneous fringe-projections having fully overlapped spatial-frequency spectra (Fig. 3). Figure 6 shows the 5 phase-shifted fringe-patterns obtained with simultaneous projection from the left and right-projectors. The 2 frequencies are $\omega_5=2\pi/5$ and $2\omega_5=4\pi/5$ radians/image. In Fig. 6 the fringes modulated by $b_1(x,y)$ travel at a phase-velocity of $\omega_5 = 2\pi/5$ radians/image in the $x$-direction, while the fringes modulated by $b_2(x,y)$ travel at a phase-velocity of $2\omega_5$ radians/image.

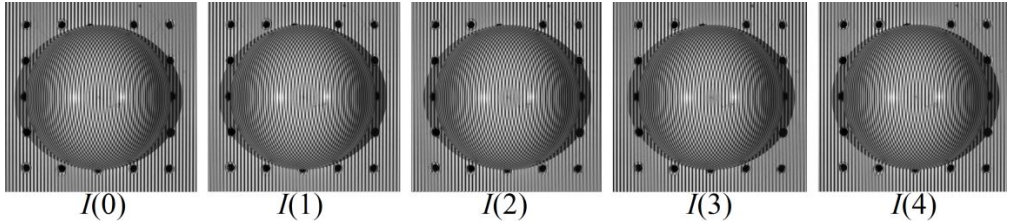

    $I(0)$        $I(1)$        $I(2)$        $I(3)$       $I(4)$

Fig. 6. The five temporal-multiplexed phase-shifted fringes obtained with simultaneous left and right projectors turned-on. The temporal-multiplexed frequencies are $\omega_5=2\pi/5$ and $2\omega_5=4\pi/5$ radians/image. Using our TV-stations analogy, each projector has its own broadcasting temporal-frequency; 2-projectors need 2-temporal-frequencies $\omega_5$ and $2\omega_5$..

The spatial-carriers $u_0x$ from the right and left projectors are equal. As seen in the previous section, the 2 PSAs tuned at $\omega_5 = 2\pi/5$ and $2\omega_5$ are:

$$\frac{b_1}{2}H_5(\omega_5)e^{igh(x,y)} = e^{-iu_0x}\left[I(0)+I(1)e^{i\omega_5}+I(2)e^{2i\omega_5}+I(3)e^{3i\omega_5}+I(4)e^{4i\omega_5}\right], \qquad (12)$$

$$\frac{b_2}{2}H_5(\omega_5)e^{-igh(x,y)} = e^{-iu_0x}\left[I(0)+I(1)e^{i2\omega_5}+I(2)e^{4i\omega_5}+I(3)e^{6i\omega_5}+I(4)e^{8i\omega_5}\right]. \qquad (13)$$

These analytic signals have different amplitudes $b_1(x,y)/2$ and $b_2(x,y)/2$ because the 2 projectors cast different shadows and glare over the digitizing solid. Finally we add these 2 signals to have a full co-phased description of the object without shadows and glare:

$$H_5(\omega_5)\left\{\frac{b_1(x,y)}{2}e^{igh(x,y)}+\left[\frac{b_2(x,y)}{2}e^{-igh(x,y)}\right]^*\right\}=\frac{H_5(\omega_5)}{2}[b_1(x,y)+b_2(x,y)]e^{igh(x,y)}. \quad (14)$$

The operator $[\cdot]^*$ stands for the complex conjugate. In co-phased profilometry the region where $b_1(x,y)=0$ is different from the region where $b_2(x,y)=0$. In other words, we obtain a well-defined analytic signal $\exp[i\,g\,h(x,y)]$ whenever $[b_1(x,y)+b_2(x,y)]\gg 0$. Therefore the co-phased sum effectively minimizes the shadow and glare regions where we would be otherwise be unable to recover a well-defined object phase.

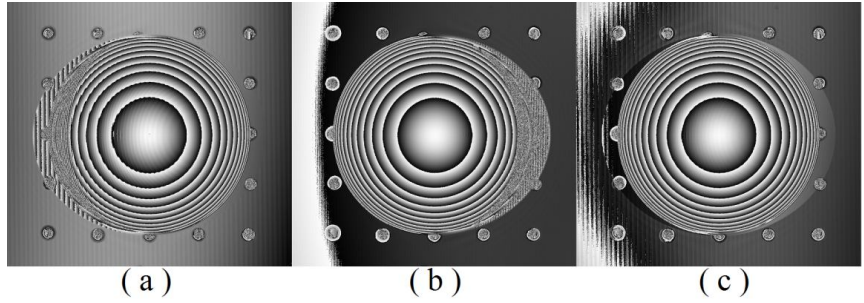

Fig. 7. Wrapped-phases estimated at: (a) $\omega_5=2\pi/5$, and at (b) $2\omega_5=4\pi/5$. Panel (c) shows their co-phased sum. Panels (a) and (b) show the shadow-regions where the modulated phase is undefined (just phase-noise is obtained). The co-phased object-phase is shown in (c); note that the self-occluding noisy shadows have been eliminated in Panel (c).

We must remark that the wrapped phases in Fig. 7(a)-7(b) are shown just for illustrative purposes. Only the co-phased wrapped sum in Fig. 7(c) is required to obtain the 3D digital rendering shown in Fig. 8.

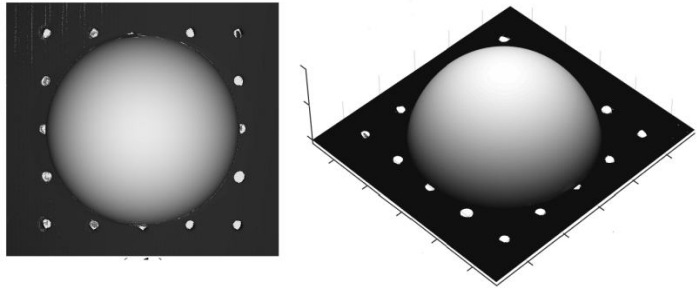

Fig. 8. The unwrapped phase shown wrapped in Fig. 7(c), and its 3D digital-rendering of the digitizing solid in Fig.1. The gray-coded phase-values have been used as texture.

## 5. Seven-step PSAs for 2 projectors and 2 frequencies multiplexed-fringes

Here we still keep 2 projectors and 2 temporal-frequencies. But now we are increasing from 5 to 7 the phase-shifted fringe-pattern number. Higher number of phase-shifted fringes improves the (S/N) and harmonic rejection.

The 7 step phase-shifted, 2 frequencies $(\omega_7, 2\omega_7)$, temporal-multiplexed fringes are,

$$I(x,y,t)=\{a+b_1\cos[g\,h+u_0 x+\omega_7 n]+b_2\cos[-g\,h+u_0 x+2\omega_7 n]\}\delta(t-n);\; \omega_7=\frac{2\pi}{7}, \quad (15)$$

being $n = \{0,1,...,6\}$. Three out-of-seven temporal-multiplexed phase-shifted fringe-patterns are shown in Fig. 9. The temporal Fourier-spectrum is equal to that shown in Fig. 4 except that in this case $(\omega_1 = \omega_7, \omega_2 = 2\omega_7)$

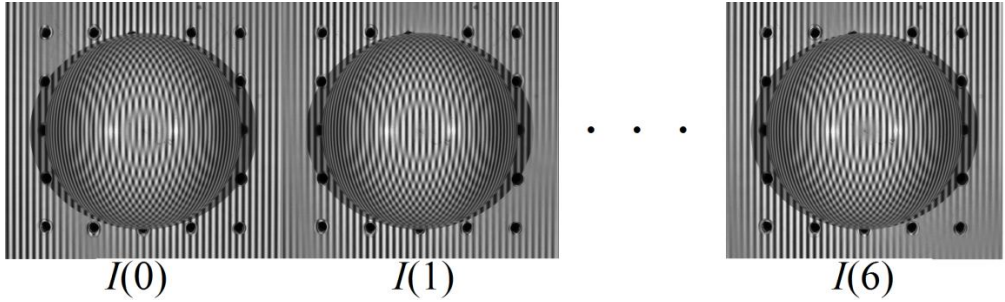

$I(0)$ $\qquad\qquad\qquad I(1) \qquad\qquad\qquad\qquad\qquad I(6)$

Fig. 9. Here we show three $\{I(0),I(1),...,I(6)\}$ out-of-seven, 2-temporal-frequencies multiplexed fringe-patterns. Using our television broadcasting analogy, each projector has its own broadcasting frequency of $\omega_7$ and $2\omega_7$ respectively. The carrier-fringes displace in the $x$-direction at a phase-velocity of $\omega_7$ and $2\omega_7$ radians/image,.

As before we need two FTFs, namely $H_7(\omega)$ and $H_7(\omega - \omega_7)$. The $H_7(\omega)$ have spectral zeroes at $\omega = \{-2\omega_7, -\omega_7, 0, 2\omega_7\}$ to isolate $(b_1/2)e^{i(gh+u_0 x)}\delta(\omega - \omega_7)$. And $H_7(\omega - \omega_7)$ filter-out the deltas at $\omega = \{-2\omega_7, -\omega_7, 0, \omega_7\}$ to keep the signal $(b_2/2)e^{i(-gh+u_0 x)}\delta(\omega - 2\omega_7)$. These two FTFs are given by:

$$H_7(\omega) = \prod_{n=0}^{5}\left[1 - e^{-i(\omega + n\omega_7)}\right], \qquad (16)$$

$$H_7(\omega - \omega_7) = \prod_{n=0}^{5}\left\{1 - e^{-i[\omega + (n-1)\omega_7]}\right\}. \qquad (17)$$

The graphs of $H_7(\omega)$ and $H_7(\omega - \omega_7)$ are shown in Fig. 10. The signal-to-noise ($S/N$) power-ration gains (with respect to the raw-data fringes) for $H_7(\omega)$ and $H_7(\omega - \omega_7)$ are [2,3],

$$\frac{|H_7(\omega_7)|^2}{\frac{1}{2\pi}\int_{-\pi}^{\pi}|H_7(\omega)|^2 d\omega} = 7; \qquad \frac{|H_7(\omega_N)|^2}{\frac{1}{2\pi}\int_{-\pi}^{\pi}|H_7(\omega - \omega_7)|^2 d\omega} = 7. \qquad (18)$$

Therefore $H_7(\omega)$ and $H_7(\omega - \omega_7)$ have both the same signal-to-noise power-ratio gain [2,3].

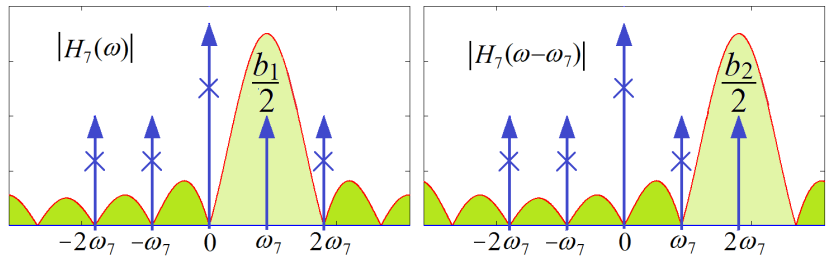

Fig. 10. Seven-step FTFs for 2 frequencies temporal-multiplexing. The FTF $H_7(\omega)$ keeps the signal located at $\delta(\omega-\omega_7)$ and rejects all other spectral components, while $H_7(\omega-\omega_7)$ preserves only the signal located at $\delta(\omega-2\omega_7)$. Vertical arrows represent Dirac deltas.

Taking the inverse Fourier transforms $F^{-1}[H_7(\omega)]$ and $F^{-1}[H_7(\omega-\omega_7)]$ the 7-step impulse responses of the PSAs are obtained as,

$$F^{-1}[H_7(\omega)] = \sum_{n=0}^{6}\delta(t-n)e^{in\omega_7}, \tag{19}$$

$$F^{-1}[H_7(\omega-\omega_7)] = \sum_{n=0}^{6}\delta(t-n)e^{2in\omega_7}. \tag{20}$$

From these impulse responses one obtains the two phase-shifting algorithms (PSAs) as [2],

$$A_1(x,y)e^{igh(x,y)} = e^{-iu_0 x}\left[\sum_{n=0}^{6}e^{in\omega_7}I(x,y,n)\right]; \tag{21}$$

$$A_2(x,y)e^{-igh(x,y)} = e^{-iu_0 x}\left[\sum_{n=0}^{6}e^{2in\omega_7}I(x,y,n)\right]. \tag{22}$$

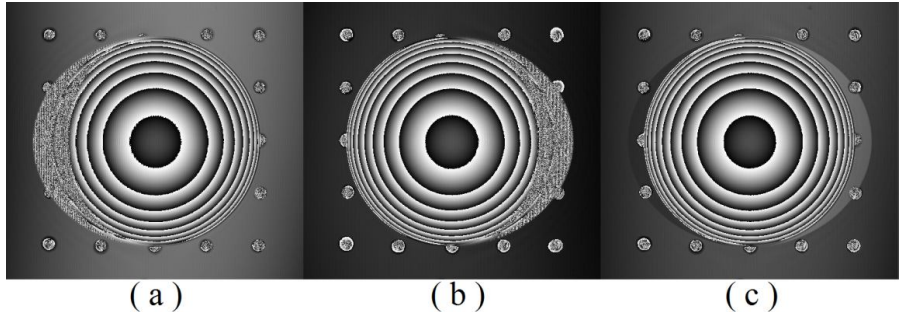

(a)          (b)          (c)

Fig. 11. (a) Shows the recovered phase at temporal-frequency $\omega_7=2\pi/7$. (b) Shows the wrapped-phase at temporal-frequency $2\omega_7$. (c) Shows the wrapped phase of the co-phased sum. In panel (c) the co-phase sum is well-defined over the entire object.

With $A_1 = (b_1/2)H_7(\omega_7)$ and $A_2 = (b_2/2)H_7(\omega_7)$. Finally, the searched co-phased analytic signal is given by the sum:

$$A(x,y)e^{igh(x,y)} = A_1(x,y)e^{igh(x,y)} + \left[A_2(x,y)e^{-igh(x,y)}\right]^* = \left[A_1(x,y) + A_2^*(x,y)\right]e^{igh(x,y)}. \tag{23}$$

Note that turning on-and-off each projector sequentially (as in [7]) would require 7 fringe-images per projector to obtain the same (S/N) gain, totalizing 14 fringe-images.

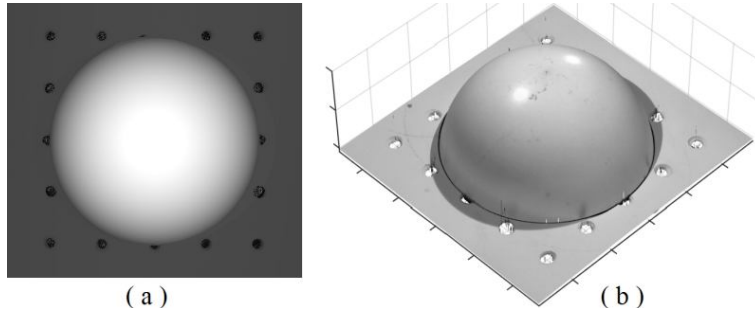

(a)          (b)

Fig. 12. Panel (a) shows the 7 step co-phased unwrapped-phase. (b) Shows a 3D digital-rendering of the co-phased, temporal-multiplexed reconstruction. In (b) we added the object image as texture and a black circle around the object for easier visualization.

The wrapped-phases for 7 temporal-multiplexed fringes are shown in Fig. 11. Figure 11(c) shows the co-phased sum of the phases at Fig. 11(a) and Fig. 11(b). The unwrapped-phase for 2 projectors and 2 temporal-frequencies multiplexing with 7 phase-steps are shown in Fig. 12.

## 6. Multiplexing of 4 fringe-projectors with 4 temporal-frequencies

As mentioned, temporal-multiplexing needs as many temporal-frequencies as simultaneous fringe-projectors are used. Therefore each projector must have its own temporal broadcasting frequency. This should not be confused with the extra phase-steps for just 2 projectors and 2 temporal-frequencies seen previously.

Figure 13 shows a schematic of the experimental set-up needed for 4 co-phased fringe-projectors aimed towards the digitizing solid. Our mathematical model for temporal-multiplexed phase-shifted fringe-images using the frequencies $\{\omega_9, 2\omega_9, 3\omega_9, 4\omega_9\}$ is,

$$I(x,y,t) = a + b_1 \cos[g\,h(x,y) + u_0 x + \omega_9 t] + b_2 \cos[-g\,h(x,y) + u_0 x + 2\omega_9 t] + \\ b_3 \cos[g\,h(x,y) + v_0 y + 3\omega_9 t] + b_4 \cos[-g\,h(x,y) + v_0 y + 4\omega_9 t]; \omega_9 = \frac{2\pi}{9}. \quad (24)$$

Being $t = \{0, 1, ..., 8\}$. Each fringe-pattern in Eq. 24 has its own broadcasting frequency $\{\omega_9, 2\omega_9, 3\omega_9, 4\omega_9\}$. So we need 4 PSAs each one tuned to $\{\omega_9, 2\omega_9, 3\omega_9, 4\omega_9\}$ respectively. Again the spatial-frequencies are equal $u_0 = v_0$. The fringes modulated by $b_1(x,y)$ and $b_2(x,y)$ move in the $x$-direction at a phase-velocity of $\omega_9$ and $2\omega_9$ radians/image respectively. The fringes modulated by $b_3(x,y)$ and $b_4(x,y)$ displace in the $y$-direction at a phase-velocity of $3\omega_9$ and $4\omega_9$ radians/image respectively.

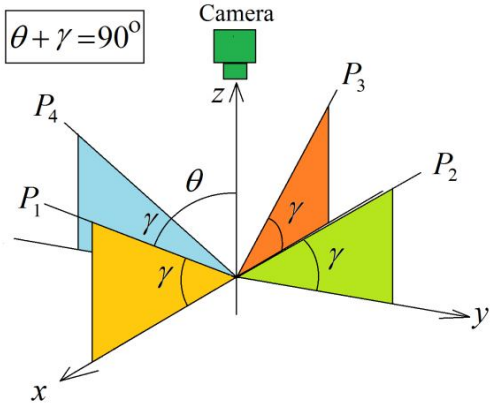

Fig. 13. Schematic of the geometry for 4 fringe-projectors aimed towards the solid located at the origin. The profilometer' sensitivity angle is $\theta$ and the spatial-frequencies for the 4 projectors ($P_1, P_2, P_3, P_4$) are the same. The fringes are projected perpendicular to the projector's directions.

The 9 temporal-multiplexed phase-shifted fringe-images (Eq. (24)) have the following spectrum (see Fig. 14),

$$I(x,y,\omega) = a(x,y)\delta(\omega) + \frac{b_1(x,y)}{2}\left[e^{i(gh+u_0 x)}\delta(\omega-\omega_9) + e^{-i(gh+u_0 x)}\delta(\omega+\omega_9)\right] +$$
$$\frac{b_2(x,y)}{2}\left[e^{i(-gh+u_0 x)}\delta(\omega-2\omega_9) + e^{-i(-gh+u_0 x)}\delta(\omega+2\omega_9)\right] +$$
$$\frac{b_3(x,y)}{2}\left[e^{i(gh+v_0 y)}\delta(\omega-3\omega_9) + e^{-i(gh+v_0 y)}\delta(\omega+3\omega_9)\right] +$$
$$\frac{b_4(x,y)}{2}\left[e^{i(-gh+v_0 y)}\delta(\omega-4\omega_9) + e^{-i(-gh+v_0 y)}\delta(\omega+4\omega_9)\right]; \omega_9 = \frac{2\pi}{9}.$$
(25)

Figure 14 shows graphically the temporal Fourier-spectra composed by 9 Dirac-deltas of the 4 temporal-multiplexed fringe-patterns in Eq. (24).

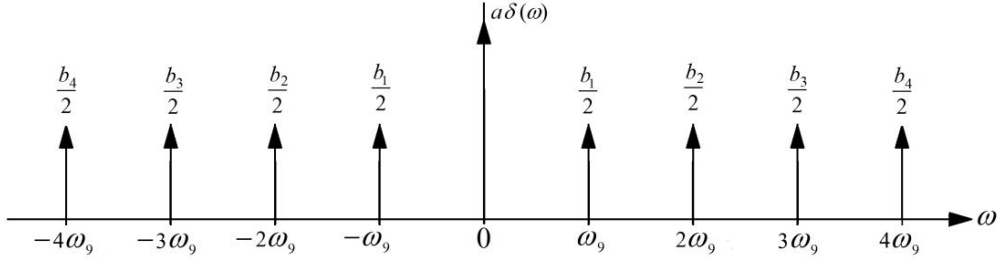

Fig. 14. Fourier spectrum of the temporal-multiplexed simultaneous-fringes using 4-temporal-frequencies $\omega_9 = 2\pi/9$, $2\omega_9$, $3\omega_9$ and $4\omega_9$. The vertical arrows represent Dirac deltas.

Figure 15 shows in a sequential way the 4 fringe projections over the spheric metallic object to clearly see the shadows and glare cast by each projector separately.

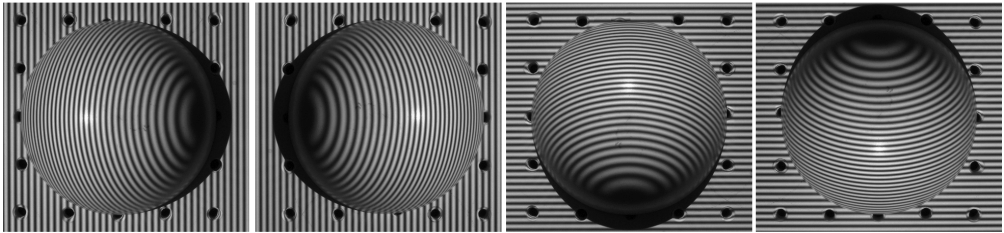

Fig. 15. Four fringe-images projected over our metallic spherical object. The fringe images are sequentially projected from the 4 different directions in Fig. 13. This is to illustrate the shadows cast and specular reflections by each projector separately.

The simultaneous projected-fringes are shown in Fig. 16 along with its overlapped Fourier spatial-spectra. As Fig. 16(a) shows, 4 fringe-projections better cover the shadows and specular reflections of the object [5,7]. The spatial-spectrum in Fig. 16(b) can only be separated by temporal-multiplexing.

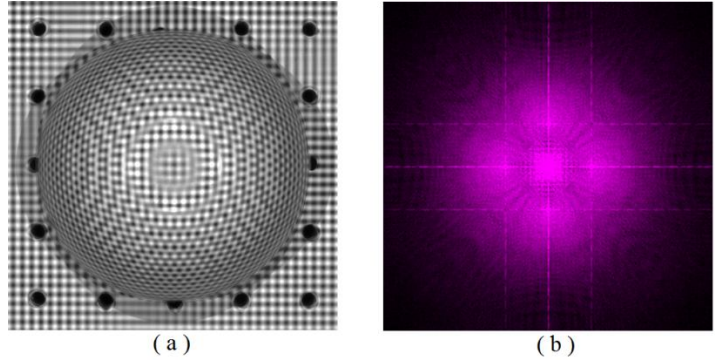

(a)   (b)

Fig. 16. Panel (a) shows one out-of-nine temporal-multiplexed phase-shifted fringes *simultaneously* illuminated by the 4 projectors in Eq. (24). The shadows and glare are better covered by 4 projectors. Panel (b) shows the overlapped Fourier spatial-spectrum which precludes their spatial-filtering. This spectrum is however well separated by the temporal-frequencies $(\omega_9, 2\omega_9, 3\omega_9, 4\omega_9)$.

We need to isolate the analytic signals at $\{\delta(\omega-\omega_9), \delta(\omega-2\omega_9), \delta(\omega-3\omega_9), \delta(\omega-4\omega_9)\}$ while filtering-out (deleting) all other spectral components using the following 4 FTFs:

$$H_9(\omega) = \prod_{n=0}^{7}\left[1 - e^{-i(\omega+n\omega_9)}\right], \qquad H_9(\omega-\omega_9) = \prod_{n=0}^{7}\left\{1 - e^{-i\left[\omega+(n-1)\omega_9\right]}\right\},$$

$$H_9(\omega-2\omega_9) = \prod_{n=0}^{7}\left\{1 - e^{-i\left[\omega+(n-2)\omega_7\right]}\right\}, \quad H_9(\omega-3\omega_9) = \prod_{n=0}^{7}\left\{1 - e^{-i\left[\omega+(n-3)\omega_9\right]}\right\}, \quad \omega_9 = \frac{2\pi}{9}. \qquad (26)$$

Figure 17 shows that $H_9(\omega)$ keeps only the signal at $\delta(\omega-\omega_9)$; $H_9(\omega-\omega_9)$ isolates the signal at $\delta(\omega-2\omega_9)$; $H_9(\omega-2\omega_9)$ keeps only the signal at $\delta(\omega-3\omega_9)$, finally $H_9(\omega-3\omega_9)$ isolates the analytic signal located at $\delta(\omega-4\omega_9)$.

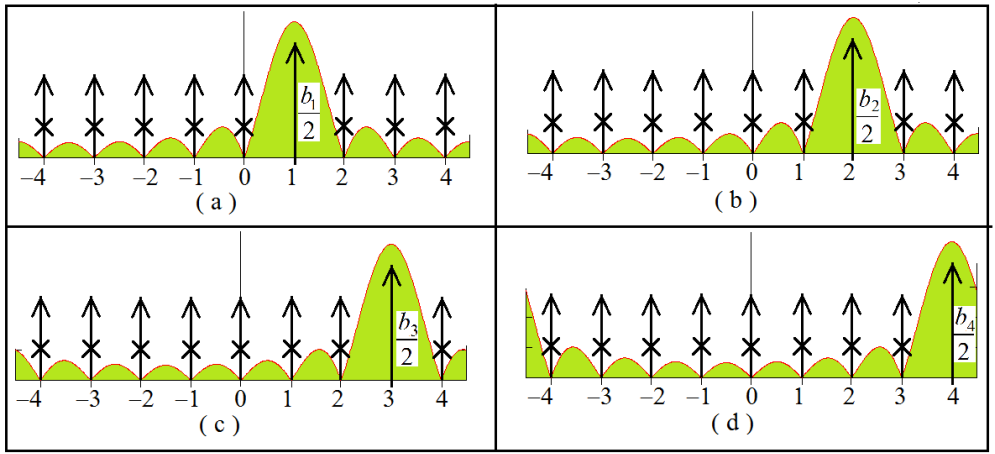

Fig. 17. Minimum 9 steps 4 FTFs for temporal-multiplexing 4 fringe-projectors. Panel (a) shows the standard 9 step least-squares FTF [2]. Panel (b), (c) and (d) show the frequency-shifted FTFs by temporal-frequencies $(2\omega_9, 3\omega_9, 4\omega_9)$ respectively. The vertical arrows represent Dirac deltas.

The (S/N) power-ratio gain of the 4 FTFs: $H_9(\omega)$, $H_9(\omega-\omega_9)$, $H_9(\omega-2\omega_9)$ and $H_9(\omega-3\omega_9)$ equals nine [2,3,7],

$$\frac{|H_9(\omega_9)|^2}{\frac{1}{2\pi}\int_{-\pi}^{\pi}|H_9(\omega)|^2 d\omega} = \frac{|H_9(\omega_9)|^2}{\frac{1}{2\pi}\int_{-\pi}^{\pi}|H_9(\omega-\omega_9)|^2 d\omega} =,...,= \frac{|H_9(\omega_9)|^2}{\frac{1}{2\pi}\int_{-\pi}^{\pi}|H_9(\omega-3\omega_9)|^2 d\omega} = 9. \quad (27)$$

The impulse responses corresponding to these 4-FTFs are,

$$F^{-1}[H_9(\omega)] = \sum_{n=0}^{8}\delta(t-n)e^{in\omega_9}, \quad F^{-1}[H_9(\omega-\omega_9)] = \sum_{n=0}^{8}\delta(t-n)e^{i2n\omega_9},$$
$$F^{-1}[H_9(\omega-2\omega_9)] = \sum_{n=0}^{8}\delta(t-n)e^{i3n\omega_9}, \quad F^{-1}[H_9(\omega-3\omega_9)] = \sum_{n=0}^{8}\delta(t-n)e^{i4n\omega_9}. \quad (28)$$

The spatial-carriers are equal $v_0 = u_0$. So the 4 searched 9-steps PSAs are given by:

$$A_1(x,y)e^{igh(x,y)} = e^{-iu_0 x}\left[\sum_{n=0}^{8}I(x,y,n)e^{in\omega_9}\right],$$
$$A_2(x,y)e^{-igh(x,y)} = e^{-iu_0 x}\left[\sum_{n=0}^{8}I(x,y,n)e^{i2n\omega_9}\right],$$
$$A_3(x,y)e^{igh(x,y)} = e^{-iv_0 y}\left[\sum_{n=0}^{8}I(x,y,n)e^{i3n\omega_9}\right], \quad (29)$$
$$A_4(x,y)e^{-igh(x,y)} = e^{-iv_0 y}\left[\sum_{n=0}^{8}I(x,y,n)e^{i4n\omega_9}\right]; \quad \omega_9 = \frac{2\pi}{9}.$$

With $A_1(x,y) = (b_1/2)H_7(\omega_7)$, $A_2(x,y) = (b_2/2)H_7(\omega_7)$ and $A_3(x,y) = (b_3/2)H_7(\omega_7)$, and $A_4(x,y) = (b_4/2)H_7(\omega_7)$. Finally the co-phased sum of these 4 analytic signals is:

$$A(x,y)e^{igh(x,y)} = \left[A_1(x,y) + A_2^*(x,y) + A_3(x,y) + A_4^*(x,y)\right]e^{igh(x,y)}. \quad (30)$$

In co-phased profilometry, one obtains a well-defined object-phase $gh(x,y)$ whenever the magnitude of $|A_1 + A_2^* + A_3 + A_4^*| >> 0$. This means that all the object self-occluding shadows and glare are covered-up by the 4 fringe-projections. Please note that we need at least 9 temporal-multiplexed fringe-patterns to perform this measurement. For comparison, co-phased fringe-projection profilometry with 4 projectors sequentially turned on-and-off would require 9x4=36 fringe-patterns to obtain the same (S/N) power-ratio gain with respect to the raw-data fringes [2,3]. Further generalization to higher number of fringe-projectors and/or higher-order PSAs is just a matter of mathematical induction.

## 7. Summary

Here we have shown how to design temporal multiplexed PSAs for co-phased fringe-projection profilometry. We have shown that turning-on all the co-phased projectors and using temporal-multiplexing one may reduce by half or more the number of fringe-images while maintaining the same (S/N) power-ratio gain. Our first experiment used 5 phase-step images with 2 simultaneous fringe-projectors and temporal-frequency ($\omega_5 = 2\pi/5, 2\omega_5$). As it is well known, the phase estimation improves by increasing the number of phase-shifted images [2]. In particular our next experiment also used 2-projectors and 2-temporal-frequencies ($\omega_7 = 2\pi/7, 2\omega_7$); we have increased to 7 fringe-patterns images. With temporal-multiplexing we are not only saving half or more fringe-patters images, but simultaneous fringe projection may be achieved much faster than turning on-and-off the fringe projectors sequentially as did before [5-7].

Finally we have generalized temporal-frequency multiplexing to 4 fringe-projectors turned-on simultaneously; each projector having a different illumination direction; each projector having also their own temporal frequency ($\omega_9 = 2\pi/9, 2\omega_9, 3\omega_9, 4\omega_9$). We have seen that we need at least 9 fringe-patterns to separate the 4 co-phased analytic-signals. Again without temporal-multiplexing one would require 4x9=36 fringe-patterns (9-fringe-images per projector) to obtain the same signal-to-noise (*S/N*) power-ratio gain.

**Acknowledgments**

The authors acknowledge Cornell University for hosting arXiv e-print repository, and the Optical Society of America for allowing the contributors to post their manuscript at arXiv.